\documentclass[12pt]{article}
\usepackage{graphics}
\begin{document}
\title{Naive dimensional analysis\\and truly strong interactions}
\author{\normalsize B.
Holdom\thanks{holdom@utcc.utoronto.ca}\\\small {\em Department of
Physics, University of Toronto}\\\small {\em Toronto, Ontario,}
M5S1A7, CANADA}\date{}\maketitle
\begin{abstract} We question whether the canonical estimate of a strong
coupling, \(\alpha \approx 4\pi \), is always appropriate for dynamical
symmetry breaking. Our discussion is motivated by the success of naive
quark models in describing low energy QCD.
\end{abstract}
\baselineskip 19pt

Naive dimensional analysis (NDA) \cite{b} has emerged as a practical
tool for estimating the dimensionless parameters appearing in a low
energy effective theory when the underlying theory is strongly-coupled.
The basic reasonable assumption is that the effective theory should reflect
the strongly-interacting nature of the underlying theory, so that loop
corrections in the effective theory should be as large as tree level effects.
This yields estimates for coefficients of operators in the effective theory.
Such estimates turn out to work rather well for low energy QCD, and
since the basic idea seems very general, it is natural to apply it to other
types of strong interactions.

Central to the application of NDA is an estimate of the size of coupling in
a strongly-coupled theory. The canonical estimate is \(\alpha \approx 4\pi
\), since it is for this coupling that all orders of the loop expansion are
expected to be of similar size. We will refer to \(\alpha \approx 4\pi \) as a
truly strong coupling, and in the absence of any small expansion
parameter, the parameters appearing in the low energy effective theory
are truly incalculable in perturbation theory. NDA is the best that can be
done in the absence of a true nonperturbative calculation. 

We will suggest that symmetry breaking physics need not always be truly
strong in the following sense. It may be that symmetry breaking physics
can at times leave a recognizable imprint on the low energy theory, an
imprint not completely obscured by nonperturbative effects and reflecting
the details of the underlying theory. We suggest that a coupling which is
not truly strong is consistent with the nonperturbative ``strongly-coupled"
effects underlying the dynamical symmetry breaking. This may be the
situation for QCD, such that the success of NDA estimates can be
reconciled with the success of naive quark model estimates in low energy
QCD. Our discussion does not challenge the basic notion of NDA.

We begin with an example which illustrates the possibility of symmetry
breaking physics without a truly strong coupling. We look at a chiral
gauge theory and compare NDA with some other standard lore, namely
that which follows from the ladder Schwinger-Dyson analysis (LSDA).
The latter relies on a one gauge boson approximation to the SD equation.
One of the most familiar chiral gauge theories is \({\it SU}(5)\) with
fermion content \(10 + \overline{5}\). At strong enough couplings it is
expected that condensates in the channels
\begin{equation}10{\times}10\rightarrow \overline{5}{\ \ \ \ \ \
}\overline{5}{\times}10\rightarrow 5\end{equation} will occur, since
these are the most attractive channels for the respective pairs of fermions
\cite{a}. LSDA gives the critical couplings needed for these condensates
as \({\alpha _{c}}=5\pi /36=.44\) and \({\alpha _{c}}=5\pi /27=.58\)
respectively. We may take the first as the critical coupling for the
symmetry breaking \({\it SU}(5)\rightarrow {\it SU}(4)\). (Whether the
second condensate is considered to arise from the \({\it SU}(5)\)
interactions or the unbroken \({\it SU}(4)\) interactions is a matter of
taste.) Far below the \({\it SU}(5)\) breaking scale and the \({\it SU}(4)\)
confining scale all degrees of freedom have decoupled from the low
energy theory except for one fermion. This is the \({\it SU}(4)\) singlet
component of the original \(\overline{5}\) fermion multiplet, the
left-handed field \({\psi _{5}}\), whose masslessness is protected by an
unbroken global symmetry
\cite{a}.

The critical couplings obtained from LSDA are well below \(\alpha
\approx 4\pi \). LSDA implies that the sum of ladder graphs in specific
attractive channels is sufficient to produce dynamical symmetry breaking.
The theory is strongly-coupled in the sense that the symmetry breaking
effect is nonperturbative; graphs at all orders are contributing. The
discrepency with NDA is simply in the value of the critical coupling in
LSDA, which is less than truly strong.

It is tempting to dismiss the naive LSDA results, based as they are on the
assumption of a constant coupling. On the other hand we may associate
this constant coupling with the value of the true running coupling when
smeared or averaged over the dominant range of momenta in loops. The
basic point then is that the LSDA result is presumably reasonable if
higher order corrections to the ladder sum are small, which may follow if
the coupling is indeed not truly strong. In fact this is consistent with
explicit calculations of next order corrections to the LSDA, where
relatively small corrections are found
\cite{c}. Thus the LSDA results appear to be self-consistent.

Dynamical symmetry breaking in theories which are not truly strong
would clearly have implications for the resulting low energy effective
theories. In the present example the exchange of a massive \({\it SU}(5)\)
gauge boson generates the following interaction.
\begin{equation}\frac {4\pi }{5}\frac {{\alpha _{c}}}{M^{2}}\sum
_{n=0}^{\infty }\overline{{\psi _{5}}}{\gamma _{\mu }}{\psi
_{5}}\left(\! \frac {{\partial^2}}{M^{2}}\!\right) ^{n}\overline{{\psi
_{5}}}\gamma ^{\mu }{\psi _{5} }{\label{a}}\end{equation} But by
NDA the effective action of the low energy theory should be of the
following form, where all dimensionless parameters appearing in the
Lagrangian are of order unity \cite{b}.
\begin{equation}S=\int d^{4}x\frac {\Lambda ^{4}}{(4\pi
)^{2}}{\cal{L}} \left(  \! \frac {4\pi {\psi _{5}}}{\Lambda ^{{\frac
{3}{2} }}}, \frac {{\partial}}{\Lambda } \!  \right)
{\label{b}}\end{equation} The set of terms in (\ref{a}) could be
consistent with (\ref{b}) only if the following relations were true. 
\begin{eqnarray}&&{\ \ \ \ \ }M{\hbox{\(
\buildrel?\over\approx \)}}
\Lambda {\label{g}}\\&&\frac {4\pi }{5}\frac {{\alpha
_{c}}}{M^{2}}{\hbox{\(\buildrel?
\over\approx \)}}\frac {(4\pi )^{2}}{\Lambda
^{2}}{\label{h}}\end{eqnarray} If we assume that (\ref{g}) is true then
the NDA estimate of the overall size of the four-fermion operators (RHS
of (\ref{h})) is 144 times larger than the LSDA estimate (the LHS)! Part
of the blame for this large discrepancy surely lies in LSDA, which ignores
the fact that the \({\it SU}(5)\) coupling is running quite quickly near the
\({\it SU}(5)\) breaking scale. Thus even though some smeared value of
the coupling appropriate for loop integrations may be close to \({\alpha
_{c}}\), the \({\alpha _{c}}\) appearing in (\ref{a}) is likely an
underestimate. On the other hand for chiral gauge theories with walking
couplings, the corresponding effect would be smaller.

The large range in possible estimates for the size of four-fermion
operators does not contradict the order-of-magnitude estimates of NDA.
But the implication from LSDA, and a coupling which is not truly strong,
is that the effect of loop corrections can presumably be quite modest. The
specific form of the four-fermion interactions in (\ref{a}) for example can
remain as a clear imprint on the low energy theory.

In the above example the mass of the heavy boson set the scale for the
momentum expansion, whereas in the next example involving a heavy
fermion the situation is a little different. Consider a standard nonlinear
\(\sigma \)-model with a triplet of massless pions coupled to a degenerate
fermion doublet of mass \(m\).
\begin{eqnarray}&&{\cal{L}}=\frac {f^{2}}{4}{\rm Tr}{D_{\mu
}}UD^{\mu }U^{{\dagger} }\mbox{} + \overline{\psi }(i{/\!
\!\!\partial} + {/
\!\!\!V} + {/\!
\!\!\!A}{\gamma _{5}})\psi  - m\overline{\psi }U\psi {\label{d}}\\&&{\
\ \ \ }{\it U}(x)\equiv e^{{ - \frac {2i\pi (x ){\gamma _{5}}}{f}}}{,\ \ \ \
}V\equiv {V_{i}}{\sigma _{i}}{,\ \ \ \ }\pi \equiv {\pi _{i}}{\sigma
_{i}}{\nonumber}\\&&{\ \ \ \ }{D_{\mu }}U\equiv {{\partial}_{\mu
}}U - i({V_{\mu }} + {A_{\mu }})U + i{\it U }({V_{\mu }} - {A_{\mu
}}){\nonumber}\end{eqnarray} The model has local \({{\it
SU}(2)_{L}}{\times}{{\it SU}(2)_{R}}\) symmetry. When used as a
model for low energy QCD, color is introduced as a global symmetry
where each quark comes in \({N_{c}}\) colors. The fermions may be
integrated out to yield the low energy theory for the massless pions. We
note that this is a theory with two independent mass terms, the \(\psi \)
mass and the \({A_{\mu }}\) mass. The pion kinetic term containing the
latter cannot be removed due to the infinite renormalization from the
fermion loop.

But we are still able to study the finite higher order terms in the
momentum expansion of the fermion loop. We consider momentum
expansions for the \({\it VV} - {\it AA}\) two-point function and the
electromagnetic pion form factor.
\begin{eqnarray}&&\int e^{{iqx}}\langle {\rm T}{V_{\mu
a}}(x){V_{\nu b}}(0) - {\rm T}{A_{\mu a}}(x){A_{\nu b}}(0)
\rangle dx=i{F_{V - A}}(q^{2})f^{2}{\delta _{{\it ab}}} ({g_{\mu \nu
}} - \frac {{q_{\mu }}{q_{\nu }}}{q^{2}})\\&&\langle {\pi
_{a}}({q_{2}}) \left|  \! {V_{b}^\mu }
 \!  \right| {\pi _{c}}({q_{1}})\rangle =i{\varepsilon _{{\it
abc}}}{F_{V\pi \pi }}(q^{2})({q_{1}^\mu } + {q_{2} ^\mu }){,\ \ \ \
}q\equiv {q_{2}} - {q_{1}}\end{eqnarray} Both form factors are
defined such that \({F_{V - A}}(0)={F_{V\pi ^{2}}}(0)=1\). The model
in fact produces the same result for the two form factors for any \(q^{2}\).
\begin{equation}{F_{V - A}}(q^{2})={F_{V\pi ^{2}}}(q^{2})=1 +
\frac {{N_{c}}}{4\pi ^{2}}\frac {m^{2}}{f^{2}}(2 - 2\sqrt{\frac
{4m^{2} - q^{2}}{q^{2}}}\arctan(
\sqrt{\frac {q^{2}}{4m^{2} - q^{2}}}))\end{equation} If we consider
the momentum expansion of this result in the form
\begin{equation}1 + \frac {{N_{c}}}{24\pi ^{2}}\frac
{q^{2}}{f^{2}}\left(\! 1 + \sum _{n=1}^{\infty }\left(\! {a_{n}}\frac
{q^{2}}{m^{2}}\!\right) ^{n}\!\right) {,\label{k}}\end{equation} we
find that the \({a_{n}}\) are significantly less than unity, as shown in Fig.
(1) for \({a_{n}}\) up to high \(n\). Instead of using \(m\) as the mass
scale in the momentum expansion, a choice \(\approx 3m\) would work
better for the lower orders, while \(2m\) (the threshold for pair
production) would work better for the very high orders. We will see that
the mass scale appearing in the momentum expansion and its connection
to the fermion mass is relevant for the next example, which is closer in
spirit to QCD.

The model here will have only one independent mass parameter so that
the coupling strength \(m/f\) is determined. The independence of \(f\) and
\(m\) in the previous model is related to the infinite renormalization,
which in turn is a consequence of the momentum independence of the
fermion mass. We will therefore look at a model which incorporates a
momentum dependent fermion mass function, which is expected in any
case when the fermion mass has a dynamical origin. The minimal model
\cite{d} which incorporates the same local chiral gauge symmetries as
(\ref{d}) is obtained by removing the \({\rm Tr}{D_{\mu }}UD^{\mu
}U^{{\dagger} }\) term in (\ref{d}) and making the following
replacement.
\begin{eqnarray}&&m\overline{\psi }U\psi \rightarrow \Sigma (x - y)
\overline{\psi }(x)\xi (x){\it X}(x, y)\xi (y)\psi (y)\\&&{\ \ \ \ }\xi
(x)^{2}\equiv {\it U}(x){,\ \ \ 
\ \ \ }{\it X}(x, y)\equiv {\rm P}e^{ \left(  \!  -  i\int _{x}^{y}{\Gamma
_{\mu }}(z)dz^{\mu } \!  \right) } {\nonumber}\\&&{\ \ \ \ }{\Gamma
_{\mu }}\equiv \frac {i}{2}(\xi ({{\partial}_{\mu }} - i{V_{\mu }} - i
{A_{\mu }}{\gamma _{5}})\xi ^{{\dagger}} + \xi
^{{\dagger}}({{\partial}_{\mu }} - i{V_{\mu }} + i{A_{\mu
}}{\gamma _{5}})\xi ) {\nonumber}\end{eqnarray}
 \(\Sigma (x - y)\) is the Fourier transform of \(\Sigma ( - p^{2})\). When
the latter function falls for large \( - p^{2}\) in the way appropriate for
QCD, the fermion loop generates a finite contribution to the \({\rm
Tr}{D_{\mu }}UD^{\mu }U^{{\dagger} }\) term.

The new quark propagator is \(i/({/\!\!\!p} - \Sigma  ( - p^{2}))\). After
expanding the quark loop in external momentum and Wick rotating, we
are left with an integral over Euclidean momentum \(P^{2}= - p^{2}{\ >\
}0\). Since the integral is dominated for \(P^{2}\approx m^{2}\) we
choose the normalization condition \(\Sigma (m^{2})=m\) in order for the
new propagator to resemble \(i/({/\!\!\!p} - m)\) in the dominant
momentum region. Thus \(m\) is again the typical mass parameter of the
underlying theory.

The proper normalization of the \({\rm Tr}{D_{\mu }}UD^{\mu
}U^{{\dagger} }\) term yields the Pagels-Stokar \cite{e} relation.  
\begin{equation}f^{2}=\frac {{N_{c}}}{4\pi ^{2}}\int
dP^{2}P^{2}\frac {\Sigma ^{2} - P^{2}\Sigma \Sigma '/2 }{(P^{2} +
\Sigma ^{2})^{2}}{\label{e}}\end{equation} A convenient choice for
\(\Sigma (P^{2})\) is
\begin{equation}\Sigma (P^{2})=\frac {(A + 1)m^{3}}{P^{2} +
Am^{2}}{\label{f}}\end{equation} and we shall present results for the
values \(A=1, 2, 3, 4\).\footnote{Of interest to the modeling of
confinement is the fact that the propagator has no pole when \(A <
4.83\).} \(A\) around 2 or 3 does a good job of reproducing the observed
parameters at order \(p^{4}\), \({L_{1}}\)--\({L_{10}}\), of low energy
QCD (after current quark masses are added to the model). Combining
(\ref{e}) and (\ref{f}) yields the following set of couplings \(m/f\), which
do not display extreme sensitivity to \(A\). 
\begin{equation}
\begin{array}{|c|c|c|c|c|c|c|c|c|c|}\hline A & 1 & 2 & 3 & 4 \\\hline m/f &
4.08 & 3.77 & 3.57 & 3.42\\\hline
\end{array}
\end{equation}

We again consider the form factors \({F_{V\pi ^{2}}}(q^{2})\) and
\({F_{V - A}}(q^{2})\), which are no longer equal. If we expand both
form factors in the way suggested by NDA,
\begin{equation}1 + \sum _{n=1}^{\infty }\left(\! {b_{n}}\frac
{{N_{c}}q^{2}}{16\pi ^{2}f^{2}}\!\right) ^{n}\end{equation} we find
the following results for the expansion parameters
\({b_{n}}\).\footnote{These numbers are converted from those listed in
\cite{f}.}
\begin{equation}
\begin{array}{|c|c|c|c|c|c|c|c|c|c|}\hline {b_{n}^{V - A}} & n=1 & 2 & 3
& 4 \\\hline A=1 & 1.42 & .92 & .81 & .80 \\ 2 & 1.16 & .78 & .66 & .63
\\ 3 & 1.05 & .73 & .61 & .56 \\ 4 & .98 & .71 & .60 & .54\\\hline
\end{array}
\end{equation}
\begin{equation}
\begin{array}{|c|c|c|c|c|c|c|c|c|c|}\hline {b_{n}^{V\pi ^{2}}} & n=1 & 2
& 3 \\\hline A=1 & .90 & .70 & .69 \\ 2 & .77 & .61 & .56 \\ 3 & .73 &
.59 & .52 \\ 4 & .70 & .58 & .52\\\hline
\end{array}
\end{equation} We see a tendency for \({b_{n}}\) to decrease for
increasing \(n\), but to the extent that the \({b_{n}}\) are close to unity,
the results are fairly consistent with the expectations of NDA for a
strongly-interacting, effective theory. Their values would suggest that the
fundamental mass scale in the underlying theory is somewhat larger than
\(4\pi f/\sqrt{{N_{c}}}\), and that the underlying theory has coupling
somewhat larger than \(4\pi /\sqrt{{N_{c}}}\). On the other hand,  \(4\pi
f/\sqrt{{N_{c}}}\) and \(4\pi /\sqrt{{N_{c}}}\) are somewhat larger than
the mass \(m\) and coupling \(m/f\) actually appearing in the underlying
theory. This is understandable from the previous example, where we
learned that the mass of the fermion in the loop is not the appropriate
scale for the momentum expansion.

We can now return to the question of how it is that both NDA and naive
quark models, such as the one above, are both successful in describing
low energy QCD. The point is that what appears to be a
strongly-interacting low energy theory is emerging from an underlying
theory with a smaller than expected coupling. In fact this coupling,
\(m/f\), in the model is very similar to the size of the gauge coupling,
\(g\approx \sqrt{4\pi }\), which emerges from the ladder
Schwinger-Dyson equation analysis of chiral symmetry breaking in QCD.
Thus the idea of symmetry breaking arising for couplings less than the
truly strong coupling \(g\approx 4\pi \) emerges again here, as it did for
our first example. Support for this idea appears in the observed form of
the chiral Lagrangian of QCD, which seems to display the imprint of
some rather trivial physics.

\section*{Acknowledgment} I thank M. Luke for a useful discussion.
This research was supported in part by the Natural Sciences and
Engineering Research Council of Canada.

\begin{center} \includegraphics{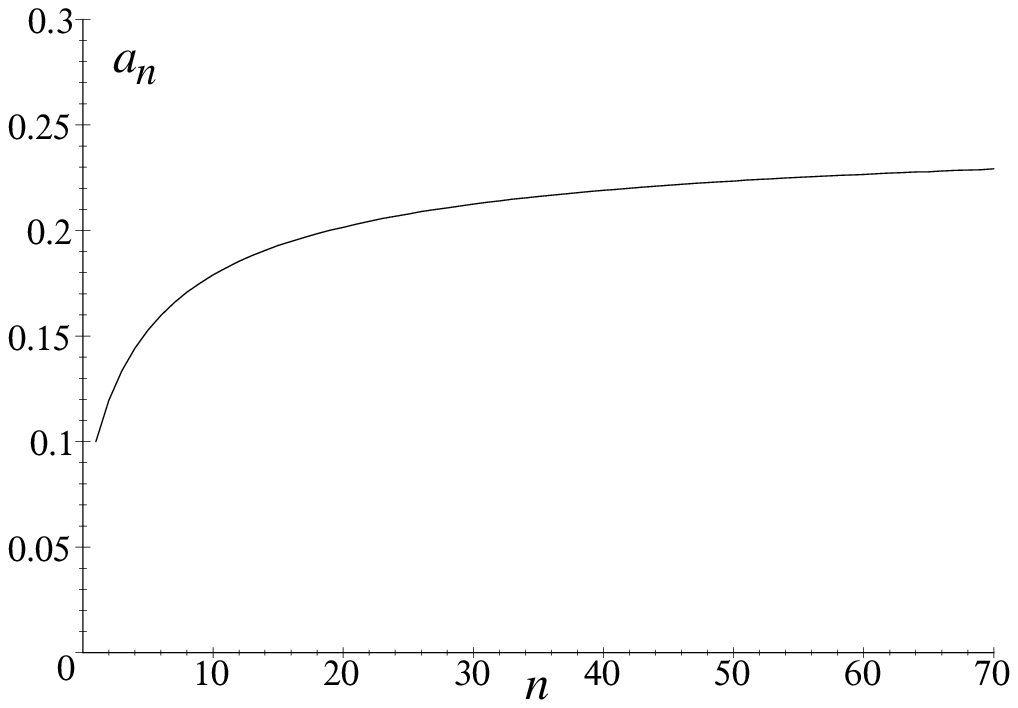}
\end{center}
\noindent Figure (1): The expansion coefficients \({a_{n}}\) in (\ref{k})
asymptotically approaching \(1/2^{2}\).

\end{document}